\documentclass{elsart}

 \usepackage{graphicx}
\usepackage{amssymb}

\begin{document}

\begin{frontmatter}



\title{Enhancement of the anomalous Hall effect and spin glass behavior in the bilayered manganite La$_{2-2x}$Sr$_{1+2x}$Mn$_2$O$_7$}


\author{Y. Hirobe\corauthref{cor1}},
\corauth[cor1]{Tel.: +81 332384108; fax: +81 332383430.}
\ead{y-hirobe@sophia.ac.jp}
\author{Y. Ashikawa},
\author{R. Kawasaki},
\author{K. Noda},
\author{D. Akahoshi}, and
\author{H. Kuwahara}

\address{Department of physics, Sophia University, Tokyo 102-8554, Japan}

\begin{abstract}
The Hall resistivity and magnetization have been investigated in the ferromagnetic state of the bilayered manganite La$_{2-2x}$Sr$_{1+2x}$Mn$_2$O$_7$ ($x$=0.36).  The Hall resistivity shows an increase in both the ordinary and anomalous Hall coefficients at low temperatures below 50K, a region in which experimental evidence for the spin glass state has been found in a low magnetic field of 1mT.  The origin of the anomalous behavior of the Hall resistivity relevant to magnetic states may lie in the intrinsic microscopic inhomogeneity in a quasi-two-dimensional electron system.
\end{abstract}

\begin{keyword}
A. Strongly correlated electron systems; E. Magnetotransport effects

\PACS 72.80.Ga, 75.40.-s, 72.20.My
\end{keyword}
\end{frontmatter}


The bilayered manganite, La$_{2-2x}$Sr$_{1+2x}$Mn$_2$O$_7$ (illustrated in the inset of Fig.1), is the $n$=2 member of the Ruddlesden-Popper series of manganites (La$_{1-x}$Sr$_{x}$)$_{n+1}$Mn$_{n}$O$_{3n+1}$.  Investigators have intensively focused on this type of manganite, possessing two-dimensional networks of MnO$_6$ octahedra, to understand the various magnetic- and/or orbital-ordered phases in its rich phase diagram \cite{Kubota,Ling}, or to clarify the collossal magnetoresistance (CMR) effect arising from strong coupling between magnetism and charge transport \cite{Moritomo,Kimura1,Kimura2,Okuda,Zhang}.  The reduction of dimensionality in manganites gives rise to anisotropic transport properties not found in the ordinary three-dimensional perovskite manganites.  For example, in bilayered manganites, the resistivity perpendicular to the MnO$_2$ plane ($\rho_{xx}^{c}$) is two orders of magnitude larger than that parallel to the plane ($\rho_{xx}^{a}$).  In addition, bilayered manganite exhibit extraordinary charge transport properties at low temperatures \cite{Moritomo,Kimura1,Kimura2,Okuda,Zhang}:  (1) a tunnel magnetoresistance (TMR) far below the ferromagnetic transition temperature $T_{\rm{C}}$, (2) a slight upturn in the temperature dependence of the longitudinal resistivity ($\rho_{xx}^{a}$ and $\rho_{xx}^{c}$) below $\sim$50K, and (3) an anomalously high resistivity of the order of 10$^{-2}{\Omega}$ cm in the metallic phase of $\rho_{xx}^{a}$, which is surprisingly high compared with that in the typical ferromagnetic metallic phase of three-dimensional perovskite manganites.  

To reveal the origin of these anomalous transport properties, researchers must perform further detailed experiments, including Hall measurements, which give information about the sign and density of mobile carriers and/or the correlation of spin and charge systems.  Chun $et$ $al$.\ have reported on low field magnetization and the Hall effect in La$_{2-2x}$Sr$_{1+2x}$Mn$_2$O$_7$ ($x$=0.40) \cite{Chun1}, where they show the occurrence of spin-glass like behavior in the magnetization measurements and an enhancement of the anomalous Hall coefficient at low temperatures.  The origin of these phenomena has been interpreted as a mixture of ferromagnetic and antiferromagnetic clusters.  According to the magnetic phase diagram for La$_{2-2x}$Sr$_{1+2x}$Mn$_2$O$_7$ reported by Kubota $et$ $al.$\cite{Kubota}, the bilayered manganite with a hole doping level of $x$=0.40 does indeed comprise a mixed state of ferromagnetic and antiferromagnetic phases, representing a canted antiferromagnetic state, a finding which also supports the above interpretation.  To remove the effect of the phase mixture, in this paper we have systematically investigated transport and magnetic properties in the $pure$ ferromagnetic metallic state without any secondary phase.  We report on both the Hall effect and the magnetization found in the ferromagnetic metallic phase of La$_{2-2x}$Sr$_{1+2x}$Mn$_2$O$_7$ with a nominal hole doping level of $x$=0.36, which gives a maximum $T_{\rm{C}}$ \cite{Kubota,Ling,Medarde}.  We have found that the reduced dimensionality and the competing interactions between ferromagnetic double-exchange and antiferromagnetic superexchange lead to a spin-glass-like magnetic state and anomalous behavior in the Hall resistivity.

A stoichiometric mixture of La$_{2}$O$_{3}$, SrCO$_{3}$, and Mn$_{3}$O$_{4}$ powder was ground and calcined three times at \(1400^\circ\)C for 24h.  A single crystalline ingot of La$_{2-2x}$Sr$_{1+2x}$Mn$_2$O$_7$ ($x$=0.36) was grown by the floating-zone method at a feeding speed of 9-12 mm/h in air.  The grown crystal was characterized at room temperature by powder x-ray diffraction measurements, which ensured that the material consisted of a single phase with a symmetry of $I4/mmm$ ($a$=0.38685(2)nm, $c$=2.02278(9)nm), without an impurity phase.  For transport measurements, we cut the crystal produced into a thin rectangular shape with typical dimensions of 2.5mm in length, 1mm in width, and 0.3mm in thickness (the $c$-axis direction).  The surface perpendicular to the $c$ axis (parallel to the MnO$_2$ bilayers) was prepared by cleaving the crystal boule.  The electrodes on the sample needed for transport measurements were made with heat-treatment-type silver paint, and copper leads were soldered onto them.  The longitudinal resistivity $\rho_{xx}^{a}$ was measured by the conventional four-probe technique with current parallel to the MnO$_{2}$ bilayers (perpendicular to the $c$ axis; see also the inset of Fig.3).  For the determination of the Hall resistivity $\rho_{xy}$, we averaged the Hall voltage by reversing the magnetic field direction at a fixed temperature to remove the offset voltage caused by the asymmetric Hall (transverse) terminals.  Measurements of AC susceptibility $\chi$ and DC magnetization $M$ were performed with Quantum Design MPMS and PPMS instruments equipped with a 1T or 9T superconducting solenoid.  We also used this apparatus to measure the time dependence of the thermoremanent magnetization.  All data for the $\chi$ and $M$ were taken during a warming process in magnetic fields after the sample had been cooled in a zero magnetic field (ZFC) or in magnetic fields (FC).

\begin{figure}[htbp]
\begin{center}
\includegraphics[width=1\linewidth,clip]{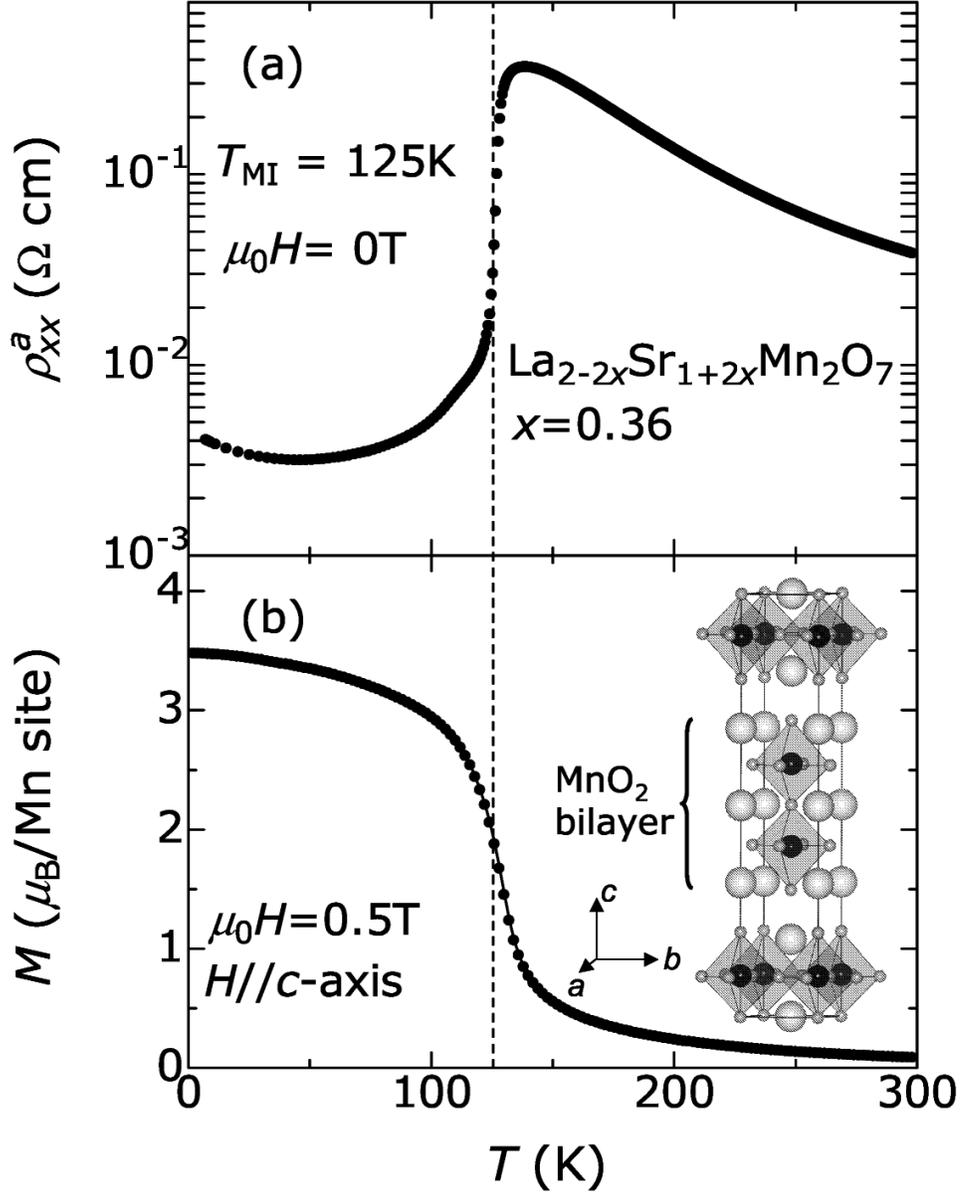}
\end{center}
\caption{Temperature dependence of (a) the in-plane ($J\bot c$) longitudinal resistivity $\rho_{xx}^{a}$, and (b) the magnetization $M$ in a magnetic field of 0.5T ($H||c$) for a La$_{2-2x}$Sr$_{1+2x}$Mn$_2$O$_7$ $x$=0.36 single crystal.  The inset in (b) shows the schematic crystal structure of La$_{2-2x}$Sr$_{1+2x}$Mn$_2$O$_7$.  The dotted line indicates the ferromagnetic (or, equivalently, metal-insulator) transition temperature $T_{\rm{C}}$.}
\label{Fig.1}
\end{figure}

Figure \ref{Fig.1}(a) shows the temperature dependence of the in-plane longitudinal resistivity $\rho_{xx}^{a}$ in a La$_{2-2x}$Sr$_{1+2x}$Mn$_2$O$_7$ ($x$=0.36) crystal, in which the current flows parallel to the conducting MnO$_{2}$ bilayer ($J$$\perp$$c$ axis).  The metal-insulator transition occurs at a temperature of $T_{\rm{MI}}$=125K, at which temperature a sharp decrease in resistivity of about two orders of magnitude is detected.  Below $T_{\rm{MI}}$, the resistivity slowly decreases down to 50K and then rises slightly.  This overall shape of the $\rho_{xx}^{a}$-$T$ curve is consistent with metallic behavior caused by the double-exchange interaction below $T_{\rm{MI}}$ in the bilayered manganites with an appropriate hole-doping level \cite{Moritomo,Kimura1,Kimura2,Okuda,Zhang}.  The observed slight upturn in the $\rho_{xx}^{a}$-$T$ curve below 50K has also been reported in the literature \cite{Moritomo,Kimura1,Kimura2,Okuda,Zhang}.  We show in Fig.\ref{Fig.1}(b) the corresponding DC magnetization for the same sample in a magnetic field of 0.5T parallel to the $c$ axis.  The magnetization steeply increases at the ferromagnetic transition temperature $T_{\rm{C}}$=125K, where the above-mentioned insulator to metal transition occurs concomitantly.  This phase transition is well known to be the transition from a paramagnetic insulating to a ferromagnetic metallic state induced by the double-exchange interaction.  The saturated magnetization of 3.6 $\mu_B$/Mn site at low temperatures reaches almost a full magnetic moment, i.e., 3.64 $\mu_B$/Mn site, a result which confirms that the sample does not contain antiferromagnetic clusters in a magnetic field of 0.5T.

\begin{figure}[htbp]
\begin{center}
\includegraphics[width=1\linewidth,clip]{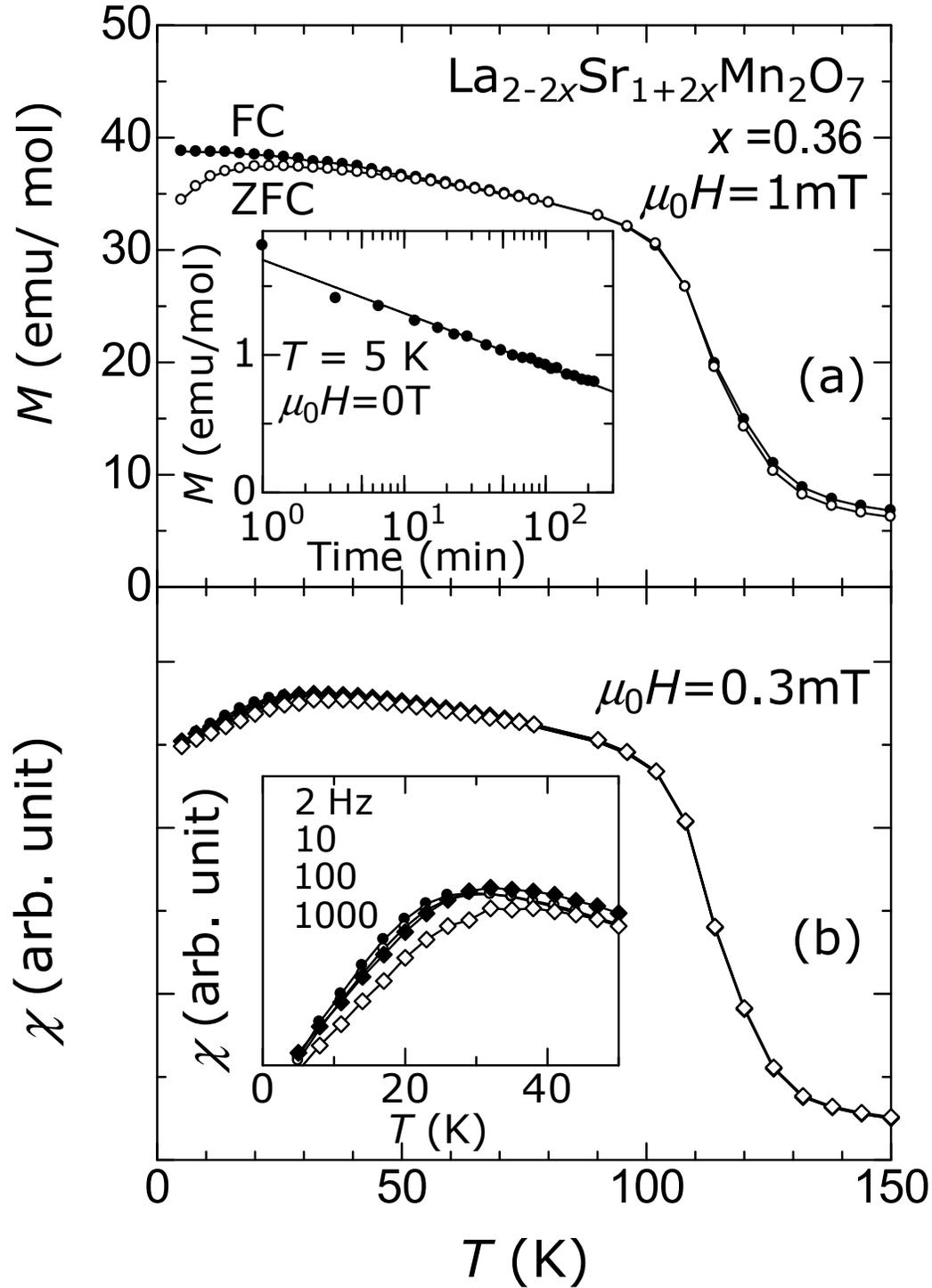}
\end{center}
\caption{(a) Temperature dependence of the field-cooled (FC, closed circles) and the zero-field-cooled (ZFC, open circles) DC magnetization for La$_{2-2x}$Sr$_{1+2x}$Mn$_2$O$_7$ $x$=0.36 single crystal in a field of 1mT.  The inset of (a) shows the logarithmic time dependence of the thermoremanent magnetization at 5K after removing $\mu_{0}$$H$=10mT (FC).  (b) Frequency-dependent AC susceptibility of the same crystal as a function of temperature.  The inset in (b) shows a magnified view below 50K.  The magnetic field was applied parallel to the $c$ axis in both cases.}
\label{Fig.2}
\end{figure}

In order to look into the spin-glass-like behavior, we have measured the DC magnetization and AC susceptibility in a small magnetic field ($\sim$1mT).  Figure \ref{Fig.2}(a) shows the temperature dependence of the DC magnetization measured in a field of 1mT after zero-field-cooling (ZFC, open circles) or field-cooling (FC, solid circles).  The following characteristic behavior was observed below $\sim$50K.  The ZFC magnetization clearly decreased with decreasing temperature; however, this result contrasted with the case of the FC process.  The observed decrease of magnetization below $\sim$50K in the ZFC process disappeared in a field of 50mT, which indicates that this ZFC-FC hysteresis below $\sim$50K easily collapses in a relatively small magnetic field.  In order to better understand the mechanism of the pronounced difference between the ZFC and FC curves, we have measured the temperature dependence of the AC susceptibility at several fixed frequencies after the ZFC process.  The real part of the AC susceptibility is shown in Fig.\ref{Fig.2}(b).  It exhibits a frequency dependence at low temperatures, and the inflection point shifts to higher temperatures as the frequency increases (see the inset in Fig.\ref{Fig.2}(b)).

We have also measured the time dependence of the thermoremanent magnetization decay at 5K in a zero field after the sample had been cooled in a field of 10mT (FC).  The result is shown in the inset of Fig.\ref{Fig.2}(a); its systematic decay conforms well to a logarithmic form.  The solid line in the inset represents: $M(t)$=$M_{0}$$-$$S$$\rm{log}_{10}$$t$, where $t$ is in minutes and the fitting parameters are $M_{0}$=1.69(2) emu/mol and $S$=0.168(5) emu/mol.  These results strongly suggest that the reentrant spin-glass-like phase exists in a nearly zero field at low temperatures below 30K.  Considering the fact that the spin-glass-like transition temperature is lower than the temperature where the slight upturn in $\rho_{xx}^{a}$ is observed, the origin of the spin-glass-like phase seems to be due to the reduction of carrier mobility caused by a weak localization effect in a low dimensional system like the present case; this effect leads to the suppression of the ferromagnetic double-exchange interaction and to the emergence of the competing antiferromagnetic superexchange one.  The spin-glass-like state, as reported in Ref.\cite{Chun1}, might be expected to be suppressed or to disappear in the $x$=0.36 sample in magnetic fields, because the spin-glass phase arises from microscopic antiferromagnetic clusters.  However, the magnetization or susceptibility results which we obtained and show in Fig.\ref{Fig.2} indicate that the spin-glass-like behavior in low magnetic fields persists even in the ferromagnetic metallic phase of the present $x$=0.36 compound.

\begin{figure}[htbp]
\begin{center}
\includegraphics[width=1\linewidth,clip]{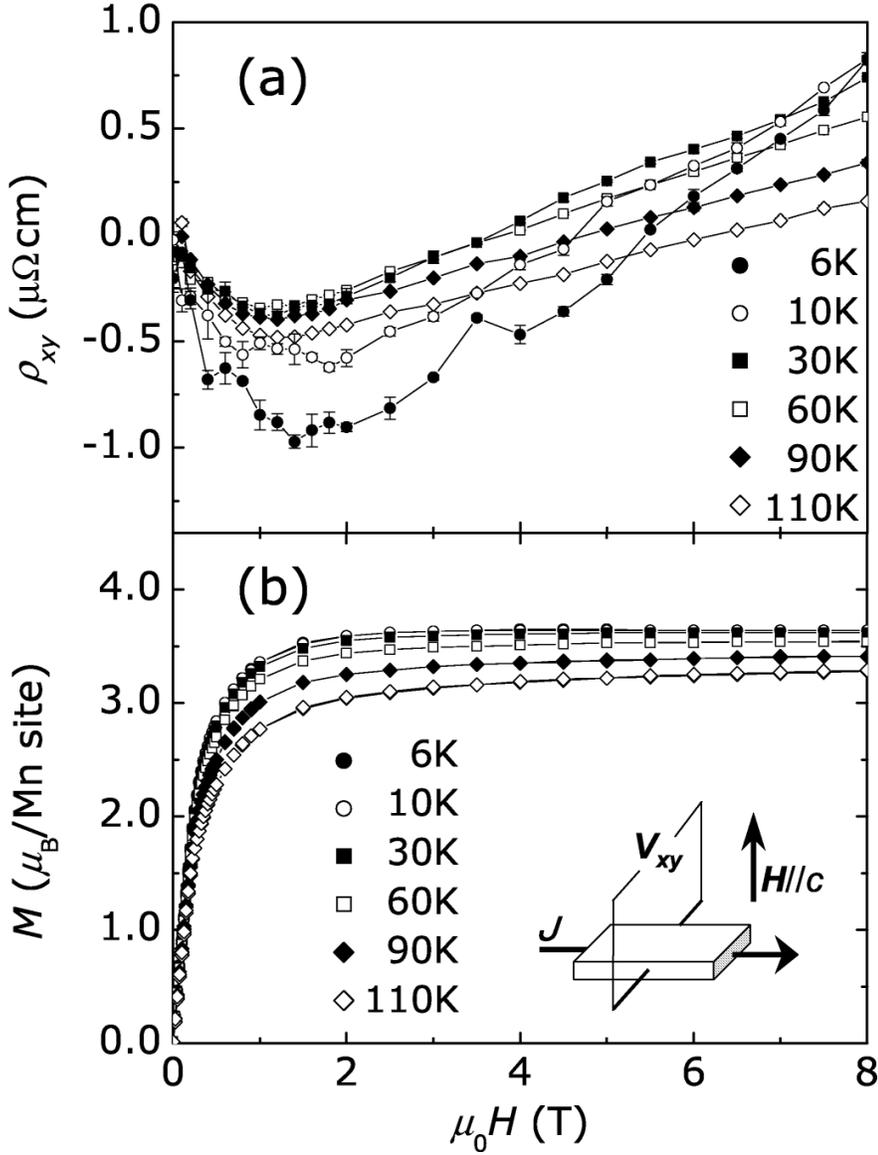}
\end{center}
\caption{Magnetic field dependence of (a) Hall resistivity $\rho_{xy}$ and (b) magnetization at several fixed temperatures below $T_{\rm{C}}$.  The lines are guide to the eyes.  The schematic of the Hall-measurement setup is shown in the inset in (b).  $M$-$H$ curves were also taken with the magnetic field parallel to the $c$ axis.}
\label{Fig.3}
\end{figure}

To obtain further insight into the relation between the magnetization and the transport properties, we have carried out Hall resistivity measurements.  This technique offers considerable promise for the understanding of the metallic phase of magnetic materials.  In order to discuss the Hall resistivity $\rho_{xy}$ quantitatively, the Hall resistivity obtained in the ferromagnetic phase was fitted to the following equation \cite{Hurd},
\begin{equation}
\rho_{xy}=R_{\rm{H}}(H+(1-N)M)+R_{\rm{S}}M
\label{eq:one}
\end{equation}
where $R_{\rm{H}}$, $R_{\rm{S}}$, $H$, $M$, and $N$ are, respectively, the ordinary Hall coefficient, anomalous Hall coefficient, external magnetic field, magnetization, and demagnetization factor determined from the sample shape ($\sim$0.8 in the present sample).  Figure \ref{Fig.3} (a) shows the Hall resistivity $\rho_{xy}$ in the $x$=0.36 crystal measured in magnetic fields increasing from 0T to 8T.  The $\rho_{xy}$ from 50K to 110K shows a behavior similar to that reported for the ferromagnetic metallic phase of three-dimensional perovskite manganites such as La$_{1-x}$Sr$_x$MnO$_3$ \cite{Asamitsu,Chun2,Chun3,Matl,Kuwahara}.  The observed negative sign of $\rho_{xy}$ in low magnetic fields indicates a negative anomalous Hall effect ($R_{\rm{S}}$$<$0), which is followed by an increase in $\rho_{xy}$ reflecting a positive ordinary Hall effect ($R_{\rm{H}}$$>$0).  Below 50K, the $\rho_{xy}$-$H$ curves appear to exhibit a nontrivial behavior; an typical example occurs at 6K, where both the ordinary and the anomalous Hall effects were observed to be enhanced, even though the $M$-$H$ curves showed normal systematic change (see Fig.\ref{Fig.3} (b)).

\begin{figure}[htbp]
\begin{center}
\includegraphics[width=1\linewidth,clip]{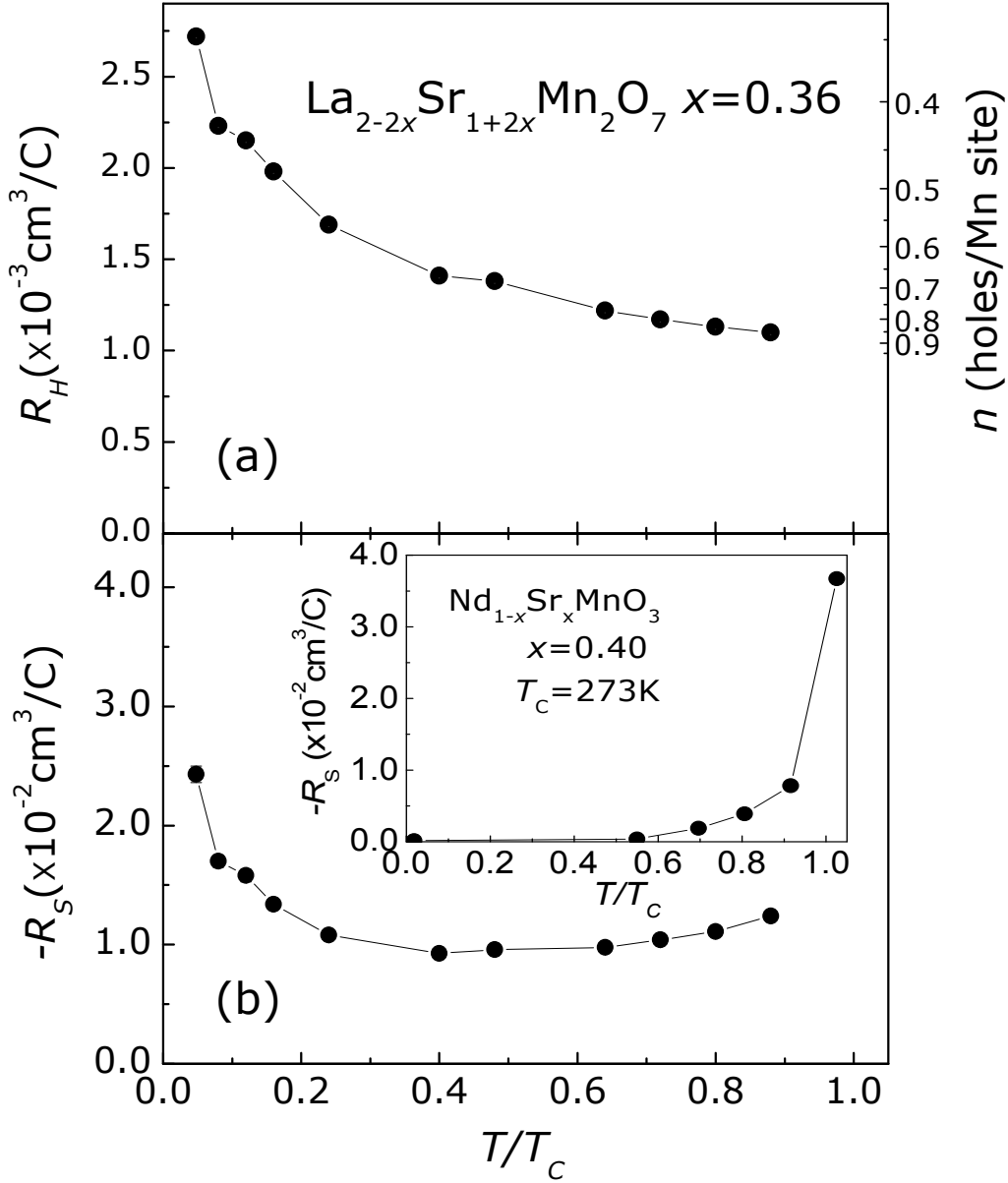}
\end{center}
\caption{(a) Temperature dependence of the ordinary Hall coefficient $R_{\rm{H}}$ for the $x$=0.36 crystal.  The effective carrier density estimated using a $n$=1/$eR_{\rm{H}}$ relation is also shown on the right ordinate.  (b) Temperature dependence of the anomalous Hall coefficient $R_{\rm{S}}$ for the same crystal.  For comparison, $R_{\rm{S}}$ for the ferromagnetic metallic phase of the three-dimensional manganite, Nd$_{1-x}$Sr$_x$MnO$_3$ $x$=0.40 \cite{Kuwahara}, is shown in the inset in (b).  The lines are guide to the eyes.}
\label{Fig.4}
\end{figure}

The temperature dependence of the fitting parameters of $R_{\rm{H}}$ and $R_{\rm{S}}$ calculated from Eq.(\ref{eq:one}), is shown in Fig.\ref{Fig.4}.  The effective carrier density $n$ shown in the right ordinate in Fig.\ref{Fig.4} (a) is estimated from a free electron model ($n$=1/$eR_{\rm{H}}$).  From the fact that the sign of $R_{\rm{H}}$ is positive, the mobile carriers in the present $x$=0.36 compound are concluded to be holes.  This result coincides with the positive sign of the carriers observed in the ferromagnetic metallic phase of three-dimensional perovskite manganites \cite{Asamitsu,Chun2,Chun3,Matl,Kuwahara}.  Concerning the temperature dependence of the effective carrier density $n$, it is reported that $n$ ($R_{\rm{H}}$) shows an almost constant value of $\sim$1 hole/Mn site in the ferromagnetic metallic phase of three-dimensional perovskite manganites \cite{Asamitsu,Chun2,Chun3,Matl,Kuwahara}, and that this value is nearly independent of temperature.  On the other hand, in the case of the present bilayered manganite, $n$ decreases ($R_{\rm{H}}$ increases) with decreasing temperature from 0.9 to 0.35 holes/Mn site: the lowest value of 0.35 is close to the nominal hole doping level of this sample ($x$=0.36), while the highest one is nearly the same as that in the ferromagnetic phase of three-dimensional manganites \cite{Asamitsu,Chun2,Chun3,Matl,Kuwahara}.\  Figure \ref{Fig.4} (b) shows the temperature dependence of $R_{\rm{S}}$ as deduced from Eq.(\ref{eq:one}).\  As the temperature decreases below $T_{\rm{C}}$, the absolute value of $R_{\rm{S}}$ falls slightly down to 50K (0.4$T_{\rm{C}}$), and then rapidly rises as the temperature sinks to a minimum (6K).  From $T_{\rm{C}}$ to 50K, the temperature dependence of $R_{\rm{S}}$ shows a similar behavior to that found in the ferromagnetic metallic phase of three-dimensional perovskite manganites; we provide these results in the inset of Fig.\ref{Fig.4} (b) \cite{Kuwahara}.  A significant increase of $R_{\rm{S}}$ below 50K is characteristic of bilayered manganites.  The temperature of 50K, where $R_{\rm{S}}$ takes a local minimum, agrees well with that observed for the local minimum of the ordinary longitudinal resistivity $\rho_{xx}^{a}$ (see Fig.\ref{Fig.1} (a)).  However, the extraordinary enhancement of $R_{\rm{S}}$ below 50K cannot be explained in terms of a conventional model based on impurity scattering, in which $R_{\rm{S}}$ can be expressed as a function of ordinary longitudinal resistivity $\rho_{xx}$ as in the following equation:
\begin{equation}
R_{\rm{S}}=(a\rho_{xx}+b\rho_{xx}^2)
\label{eq:two}
\end{equation}
where $a$ and $b$ are temperature- and magnetic-field-independent constants.  The first term in Eq.(\ref{eq:two}) describes the skew scattering \cite{Smit} and the second term from the side-jump \cite{Berger}.

To discuss the nontrivial enhancement of the anomalous Hall coefficient $R_{\rm{S}}$ at low temperatures, we return to Fig.\ref{Fig.3}.\  The anomalous Hall term $R_{\rm{S}}$ is characterized by a decrease of $\rho_{xy}$ with a negative slope observed in magnetic fields up to 1T.  This result suggests that the enhancement of $R_{\rm{S}}$ at low temperatures is related to the magnetization up to 1T.  Therefore, the enhancement of $R_{\rm{S}}$ cannot be directly attributed to the spin-glass-like phase evidenced in Fig.\ref{Fig.2}, because this phase is observed only in magnetic fields less than 50mT, fields that are too small to cause the anomalous reduction in $\rho_{xy}$ in fields up to 1T.  The origin of the glassy behavior and the enhancement of $R_{\rm{S}}$ at low temperatures is not yet clear.  A possible interpretation of these phenomena might be an intrinsic phase inhomogeneity of the electron system due to the reduction of bandwidth; this reduction, in turn, might be caused by the diminished dimensionality and quenched disorder or by a local structural distortion originating in the variance of the ionic radii of La and Sr ions in manganites \cite{Tomioka}.  The slight upturn in $\rho_{xx}^{a}$ is interpreted as a weak localization effect in the low-dimensional electron system.  This localization effect may persist up to a magnetic field as high as 7T \cite{Okuda}, and is one of the possible origin of the observed enhancement of $R_{\rm{S}}$ at low temperatures.

In summary, a spin-glass like behavior in low magnetic fields and an extraordinary increase of the anomalous Hall effect are observed below 50K in the bilayered manganites La$_{2-2x}$Sr$_{1+2x}$Mn$_2$O$_7$ with a nominal doping level of $x$=0.36.  This spin-glass like behavior persists even at the doping level $x$=0.36 with a maximum $T_{\rm{C}}$, indicating that the glass-like behavior is a common feature of the ferromagnetic metallic phase of bilayered manganites.  The origin of the extraordinary increase in the anomalous Hall effect may lie in a reduction of carrier mobility due to the reduction of dimensionality which, in turn, leads to an intrinsic microscopic phase-inhomogeneity of the electron system.  Further detailed investigations, including for example, Hall measurements taken in the ferromagnetic metallic phase with a finely-controlled bandwidth, are necessary to understand the origin of the anomalous magneto-transport behavior in bilayered manganites.

\newpage


\begin{thebibliography}{99}
\bibitem{Kubota} M. Kubota, H. Fujioka, K. Hirota, K. Ohoyama, Y. Moritomo, H. Yoshizawa, and Y. Endoh, J. Phys. Soc. Jpn. {\bf 69}, 1606 (2000).
\bibitem{Ling} C. D. Ling, J. E. Millburn, J. F. Mitchell, D. N. Argyriou, and J. Linton, Phys. Rev. {\bf B 62}, 15096 (2000).
\bibitem{Moritomo} Y. Moritomo, A. Asamitsu, H. Kuwahara, and Y. Tokura, Nature (London) {\bf 380}, 141 (1996).
\bibitem{Kimura1} T. Kimura, Y. Tomioka, H. Kuwahara, A. Asamitsu, M. Tamura, and Y. Tokura, Science {\bf 274}, 1698 (1996).
\bibitem{Kimura2} T. Kimura, Y. Tomioka, A. Asamitsu, and Y. Tokura, Phys. Rev. Lett. {\bf 81}, 5920 (1998).
\bibitem{Okuda} T. Okuda, T. Kimura, and Y. Tokura, Phys. Rev. {\bf B 60}, 3370 (1999).
\bibitem{Zhang} C. L. Zhang, X. J. Chen, C. C. Almasan, J. S. Gardner, and J. L. Sarrao, Phys. Rev. {\bf B 65}, 134439 (2002).
\bibitem{Chun1} S. H. Chun, Y. Lyanda-Geller, M. B. Salamon, R. Suryanarayanan, G. Dhalenne, and A. Revcolevschi, J. Appl. Phys. {\bf 90}, 6307 (2001).
\bibitem{Medarde} M. Medarde, J. F. Mitchell, J. E. Millburn, S. Short, and J. D. Jorgensen, Phys. Rev. Lett. {\bf 83}, 1223 (1999).
\bibitem{Hurd} C. M. Hurd, {\it The Hall Effect in Metals and Alloys} (Plenum, New York, 1972).
\bibitem{Asamitsu} A. Asamitsu and Y. Tokura, Phys. Rev. {\bf B 58}, 47 (1998).
\bibitem{Chun2} S. H. Chun, M. B. Salamon, Y. Tomioka, and Y. Tokura, Phys. Rev. {\bf B 61}, 9225 (2000).
\bibitem{Chun3} S. H. Chun, M. B. Salamon, and P. D. Han, Phys. Rev. {\bf B 59}, 11155 (1999).
\bibitem{Matl} P. Matl, N. P. Ong, Y. F. Yan, Y. Q. Li, D. Studebaker, T. Baum, and G. Doubinina, Phys. Rev. {\bf B 57}, 10248 (1998).
\bibitem{Kuwahara} H. Kuwahara, R. Kawasaki, Y. Hirobe, S. Kodama, and A. Kakishima, J. Appl. Phys. {\bf 93}, 7367 (2003).
\bibitem{Smit} J. Smit, Physica {\bf 21}, 877 (1955).; $ibid$. {\bf 24}, 39 (1958).
\bibitem{Berger} L. Berger, Phys. Rev. {\bf B 2}, 4559 (1970).
\bibitem{Tomioka} Y. Tomioka and Y. Tokura, Phys. Rev. {\bf B 70}, 14432 (2004).



\end{thebibliography}

\newpage

\end{document}